\documentclass{PoS}

\title{Scaling  quark gluon plasma by HBT interferometry with lepton pairs
}

\ShortTitle{Scaling QGP by lepton-pair interferometry}

\author{\speaker{Payal Mohanty} and Jan-e Alam\\
        Variable Energy Cyclotron Centre, 1/AF, Bidhannagar, Kolkata- 700064, India\\
        E-mail: \email{payal@vecc.gov.in, jane@vecc.gov.in}}

\abstract{We study the intensity interferometry with lepton pairs for 
nuclear collisions at RHIC and LHC energies.  
It is argued that the invariant mass dependence of HBT radii 
extracted from the correlation functions of dilepton pairs  can be used 
as an efficient tool to scale the size and life time
of the quark gluon plasma expected to be formed in nuclear collisions 
at RHIC and LHC. Quantitatively  different magnitudes of HBT radii 
are obtained at RHIC and LHC indicating stronger radial flow at LHC.
}

\FullConference{The Seventh Workshop on Particle Correlations and Femtoscopy\\
		 September 20 - 24 2011\\
		 University of Tokyo, Japan}

\begin{document}

\section{Introduction} 
The nuclear collisions at ultra-relativistic energies have been 
effected at Relativistic 
Heavy Ion Collider (RHIC) and Large Hadron Collider (LHC)  to 
create and study the properties of thermal phase of
quarks and gluons which is called quark gluon plasma (QGP).
Several observables have been proposed to study the 
properties of QGP~\cite{hwa} - among those the electromagnetic probes -
photons and dileptons are advantageous because (i) they are produced 
at every space-time point of the QGP (ii) they leave the system without 
rescatterings and hence can deliver the information of the source point 
without any distortion (~\cite{emprobes} for review).  

\par
It is well know that the electromagnetic probes can be used as 
thermometer~\cite{LM} as well as a flowmeter too~\cite{PM_HBT,PM_flow}.
The transverse  momentum ($k_T$) distribution of
photons and dileptons 
reflect the temperature of the source. Because their productions
from  a thermal source depend on the temperature ($T$)
of the bath through the thermal phase space 
factors of the participants of the reaction that produces the photons and dileptons.
However, the thermal phase space factor may be changed
by  several factors -  {\it e.g.} the transverse
kick due to flow received by  low $k_T$ photons from the low 
temperature hadronic phase will mingle with 
the high $k_T$ photons from the partonic phase, making
the task of detecting and characterizing the QGP difficult.
For dilepton the situation is, however, different because in this
case we have two kinematic variables to describe the spectra 
- out of these two, the
$k_T$ spectra of lepton pairs is affected 
but the $k_T$ integrated invariant mass ($M$) spectra
is unaltered by the  flow.   Moreover, the $M$ spectra of thermal dileptons
is dominated by the late hadronic phase for $M\lesssim m_\rho$ and by the early QGP 
phase for $M>m_\phi$, where $m_\rho$ and $m_\phi$ are $\rho$ and $\phi$ mesons 
mass respectively.  Therefore, the $M$ spectra of dileptons can be used as a
clock for heavy ion collisions - large $M$ corresponds to early
time and small $M (\sim m_\rho$)
correlate to large time. 
This  suggests that a judicious choice of $k_T$ and $M$ windows will be very useful to 
map the temperature and flow of the evolving matter.
%%%%%%%%%%%%%%%%%%%%%%%%%%%%%%%%%%%%%%%%%%%%%%%%%%%%%%%%%%%%%%%%%%
\section{Bose Einstein Correlation Function for Dileptons}
%%%%%%%%%%%%%%%%%%%%%%%%%%%%%%%%%%%%%%%%%%%%%%%%%%%%%%%%%%%%%%%%%%
Experimental measurements of two-particle Hanbury-Brown Twiss (HBT) 
intensity interferometry has been established as an useful tool 
to study the space-time evolution of 
the heavy-ion reaction~\cite{hb3}. The HBT interferometry with
lepton pairs (or with virtual photons which differ form lepton
pairs by a known factor) proceeds with the 
computation of the Bose-Einstein correlation function for two 
identical lepton pairs defined as,
\begin{equation}
C_{2}(\vec{k_{1}}, \vec{k_{2}}) = \frac{P_{2}(\vec{k_{1}}, \vec{k_{2}})}
{P_{1}(\vec{k_{1}}) P_{1}(\vec{k_{2}})} 
\label{eq1}
\end{equation}
where $\vec{k_i}$ is the three momentum of the pair, $i$ and 
$P_{1}(\vec{k_{i}})$
and $P_{2}(\vec{k_{1}}, \vec{k_{2}})$ represent 
the one- and two- particle inclusive
lepton pair spectra respectively
 and is expressed as follows: 
%%%%%%%%%%%%%%% Eqn. 5.30 %%%%%%%%%%%%%%%%%
\begin{equation}
P_{1}(\vec{k}) = \int d^{4}x~\omega (x,k)
\label{eqn_P1}
\end{equation}
%%%%%%%%%%%%%%%%%%%%%%%%%%%%%%%%%%%%%%%%%
and
%%%%%%%%%%%%%%% Eqn. 5.31 %%%%%%%%%%%%%%%%%
\begin{eqnarray}
P_{2}(\vec{k_{1}},\vec{k_{2}})&=& 
P_{1}(\vec{k_{1}})P_{1}(\vec{k_{2}})
+\frac{\lambda}{3}\int d^{4}x_{1} d^{4}x_{2} ~\omega (x_{1},K)
\omega (x_{2},K)~\cos(\Delta x^{\mu} q_{\mu}) 
\label{eqn_P2}
\end{eqnarray}
%%%%%%%%%%%%%%%%%%%%%%%%%%%%%%%%%%%%%%%%%

where $K=(k_1+k_2)/2$, $\Delta k_\mu=k_{1\mu}-k_{2\mu}=q_\mu$, 
$\Delta x_\mu=x_{1\mu}-x_{2\mu}$. $x_{i\mu}$ and $k_{i\mu}$ are the
%$x_i= (\tau_i\cosh{\eta_i},r_i\cos{\theta_i},r_i\sin{\theta_i},\tau_i\sinh{\eta_i})$ and 
%$k_i=(M_{iT}\cosh{y_i},k_{iT}\cos{\psi_i},k_{iT}\sin{\psi_i},M_{iT}\sinh{y_i})$ are the 
four co-ordinates for position and momentum variables 
respectively. $\omega(x,k)$ is the source function 
related to the thermal emission of  lepton pairs
per unit four volume, expressed as follows:
%%%%%%%%%%%%%%% Eqn. 5.32 %%%%%%%%%%%%%%%%%
\begin{equation}
\omega(x,k)=\int_{M_1^2}^{M_2^2}\,dM^2\,\frac{dR}{dM^2d^2k_Tdy}
\label{omega}
\end{equation}
%%%%%%%%%%%%%%%%%%%%%%%%%%%%%%%%%%%%%%%%%
The inclusion of the spin of the  lepton pairs (corresponding to the 
spin of virtual photon, which is 3) 
will reduce the value of $C_2-1$ by 1/3. 
The correlation functions can be evaluated  by using 
Eqs.~\ref{eq1}, \ref{eqn_P1}, ~\ref{eqn_P2} and ~\ref{omega}
for different average mass windows,
$\langle M\rangle$ ( $ \equiv M_{e^{+}e^{-}}$)= $(M_1+M_2)/2$.
The leading order process through which lepton pairs are produced in QGP 
is $q\bar{q}\rightarrow l^+l^-$~\cite{qqpair}. For the low $M$ dilepton 
production from the hadronic phase the decays of the light vector mesons 
$\rho, \omega$ and $\phi$ have been considered
including the continuum~\cite{emprobes,shu}.  Since the continuum 
part of the vector meson spectral functions are included,   
the processes like four pions annihilation~\cite{4pi} are excluded to avoid 
double counting. 
%%%%%%%%%%%%%%%%%%%%%%%%%%%%%%%%%%%%%%%%%%%%%%%%%%%%%%%%%%%%%%
%\section{Space-time evolution} %%%%%%%%%%%%%%%%%
%%%%%%%%%%%%%%%%%%%%% Table 1 %%%%%%%%%%%%%%%%%%%%%%%%%%%%%%% 
%\renewcommand{\arraystretch}{1.5}
%\vskip 0.2in
\begin{table}[h]
\begin{center}
\caption{Values of the various parameters used in the relativistic hydrodynamical calculations.}
\label{initialconditions}
\begin{tabular}{|c|c|c|}
\hline
%& & $R_{inv}$ & $R_{out}$ & $R_{side}$ & $R_{long}$  \\
%&  &   &  &     &   \\
%\hline
%Initial Temperature ($T_{i}$) & 290 MeV\\
$Input$ & RHIC & LHC\\
\hline
$N_f$   & 2.5 & 3\\
\hline
$T_{i}$ & 290 MeV & 640 MeV\\
\hline
%Initial time ($\tau_{i}$)& 0.6 fm\\
$\tau_{i}$& 0.6 fm & 0.1 fm\\
\hline
%Critical Temperature ($T_{c}$) & 175 MeV\\
$T_{c}$ & 175 MeV & 175 MeV\\
\hline
%Chemical Freeze-out Temperature ($T_{ch}$) & 170 MeV\\
$T_{ch}$ & 170 MeV & 170 MeV\\
\hline
%Kinetic Freeze-out Temperature ($T_{fo}$) & 120 MeV\\
$T_{fo}$ & 120 MeV & 120 MeV\\
\hline
EoS & 2+1 Lattice QCD & 2+1 Lattice QCD\\
\hline
\end{tabular}
\end{center}
\end{table}

For the space time evolution of the system relativistic hydrodynamical
model with cylindrical symmetry~\cite{hvg} and  boost invariance along
the longitudinal direction~\cite{jdb} has been used. The initial 
temperature ($T_{i}$) and proper thermalization time 
($\tau_{i}$) of the system 
is constrained by the hadronic multiplicity ($dN/dy$) 
through the relation  $dN/dy\sim T_i^3\tau_i$.
The equation of state (EoS) which governs the rate of expansion/cooling
has been taken from the lattice QCD calculations~\cite{MILC}.
The chemical ($T_{ch}$) and kinetic ($T_{fo}$) freeze-out temperatures 
are fixed by the particle ratios and the slope of the $k_T$ 
spectra of hadron~\cite{hirano}. The values of these parameters are
displayed in Table~\ref{initialconditions}.

With all the ingredients mentioned above 
we evaluate the correlation function $C_2$
for different invariant mass windows  for 0-5\% Au+Au collisions with 0-5\% 
centrality at $\sqrt{s_{NN}}$ = 200 GeV  and  Pb+Pb collisions at 
$\sqrt{s_{NN}}$ = 2.76 TeV  as a function of $q_{side}$ and 
$q_{out}$ which are related to transverse momentum of 
individual pair as follows (see~\cite{PM_HBT} for details):
\begin{eqnarray} 
 q_{out}&=&(k_{1T}^2-k_{2T}^2)/f(k_{1T},k_{2T})\nonumber\\
q_{side}&=&(2 k_{1T}k_{2T}\sqrt{1-\cos^{2}(\psi_{1}-\psi_{2})})/
{f(k_{1T},k_{2T})}
\end{eqnarray}
where $f(k_{1T},k_{2T})=
{\sqrt{k_{1T}^2+k_{2T}^2+2 k_{1T}k_{2T}\cos(\psi_1-\psi_2)}}$.

%******************************
\section{The HBT Radii}
\begin{figure}[h]
\begin{center}
\includegraphics[scale=0.4]{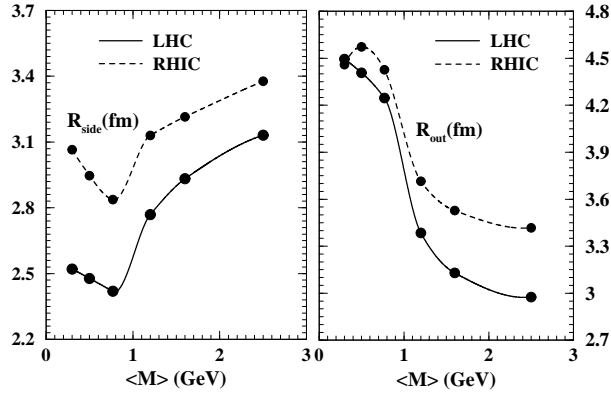}
\caption{The variation of $R_{side}$(left panel) and $R_{out}$(right panel) 
with $\langle M\rangle$ for RHIC (dashed line) and LHC (solid line) energies.
$R_{side}$ is evaluated with $k_{1T}=k_{2T}=1$ GeV and $\psi_2=0$ and
$R_{out}$ for $\psi_1=\psi_2=0$ and $k_{1T}=1$ GeV.
}
\label{fig1}
\end{center}
\end{figure}

%******************************
The dimension of the source can be 
obtained by parameterizing the calculated correlation function ($C_2$)
of the lepton pairs by the empirical (Gaussian) form:
%%%%%%%%%%%%%%%%%%%%%%%%%%%%%%%%%%%%%%%%%%
\begin{equation}
C_2=1+\frac{\lambda}{3}\exp(-R^2_{i}q^2_{i}).
\label{ceqn} 
\end{equation}
%%%%%%%%%%%%%%%%%%%%%%%%%%%%%%%%%%%%%%%%%%
where the subscript $i$ stand for $out$ and $side$ and 
$\lambda$ represents the degree of chaoticity of the 
source. The deviation of $\lambda$ from 1 will indicate 
the presence of non-chaotic sources. We have evaluated 
the $C_2$ for $\langle M \rangle=$0.3, 
0.5, 0.7, 1.2, 1.6 and 2.5 GeV and  the HBT radii are extracted 
by using Eq.~\ref{ceqn}.  The $R_{\mathrm side}$ 
scales the transverse dimension  
and the $R_{\mathrm out}$  measures both the 
transverse size and duration of particle emission~\cite{hb3,uaw}. 
The extracted $R_{\mathrm side}$ and $R_{\mathrm out}$ 
(using Eq.~\ref{ceqn}) for different 
$\langle M\rangle$ are shown in  Fig.~\ref{fig1} both for RHIC and LHC energies. 

%%%%%%%%%%%%%% Fig. 1 %%%%%%%%%%%%%%%%%%%%%%%%%%%%%

%%%%%%%%%% End of Fig.1 %%%%%%%%%%%%%%%%%%%%%%%%%%%%%
In Fig ~\ref{fig1} (left panel) we display the variation of 
$R_{side}$ with $M$.   It can be shown that the $R_{side}$ 
is related to the collective motion of the system through the relation:
$R_{side}\sim 1/(1+E_{\mathrm collective}/E_{\mathrm thermal})$. 
With the onset of transverse expansion a rarefaction 
wave moves toward the center of the cylindrical geometry -  as a consequence 
the transverse dimension of the emission zone reduces with time. Therefore, the 
size of the emission region is larger at early times and smaller at late times. 
The high $\langle M\rangle$ regions are dominated by the early partonic phase 
where the collective flow ~\cite{PM_flow} is not developed fully {\it i.e.} 
the ratio of collective to thermal energy is small hence show larger 
$R_{\mathrm side}$ of the source.  The lepton pairs with $M\sim m_\rho$ 
are emitted from the late hadronic phase where the size of the emission zone 
is smaller due to larger collective flow effects giving rise to a smaller 
$R_{\mathrm side}$. The ratio,
$E_{\mathrm collective}/E_{\mathrm thermal}$ 
is quite large for $M\sim m_\rho$ and hence $R_{side}$ is
small  which is reflected as a dip in 
the variation of $R_{\mathrm side}$ 
with $\langle M\rangle$ around the 
$\rho$-mass region (Fig.~\ref{fig1} left panel). 
Thus the variation of $R_{\mathrm side}$ with $M$  can be used as a yardstick 
to scale the size of the evolving matter.
The change of $R_{side}$ 
with $\langle M\rangle$ for RHIC and LHC is qualitatively similar
but quantitatively different. The smaller values of $R_{side}$ 
for LHC is due to the larger radial expansion which can be understood from the
fact that the  quantity $E_{\mathrm collective}/E_{\mathrm thermal}$ 
is larger at LHC than RHIC. 

\par
The $R_{\mathrm out}$ probes both the transverse size and the 
duration of emission.  The large $M$ regions are
populated by lepton pairs from the partonic phase where the
effect of flow is low (size is large) but the duration of emission is 
small - resulting in a small values of $R_{\mathrm out}$. 
Lepton pair with $M\sim m_\rho$ suffers from larger flow effects 
which should have resulted in a minimum as in $R_{\mathrm side}$ in
this $M$ region. However, $R_{\mathrm out}$ probes the duration 
of emission too, which is large for hadronic phase because of  
the  slower expansion due to softer EoS used in the present work for
the hadronic phase.  The larger duration compensates
the reduction of $R_{\mathrm out}$ 
due to flow 
%($M\sim m_\rho$ domain)
resulting is a bump in $R_{\mathrm out}$ for  $M\sim m_\rho$
(Fig.~\ref{fig1} right panel). 
The $R_{\mathrm out}$ at LHC is smaller than RHIC because
the larger flow (corresponds to smaller size) at LHC compensates
other factor (like duration of emission) which has an enhancing effect 
on $R_{\mathrm out}$.

\par
The  $R_{\mathrm out}$ and $R_{\mathrm side}$ are the 
measures of the average size of the system~\cite{rischke} and
has some dependence on space-time evolution models that is used.
However, in the
ratio, $R_{\mathrm out}/R_{\mathrm side}$  some of the uncertainties
associated with this modeling
get canceled out. The ratio, $R_{\mathrm out}/R_{\mathrm side}$
gives the duration of particle emission~\cite{hermm,chappm,hb11}
for various domains of $\langle M\rangle$ corresponding to different 
time slices.

%%%%%%%%%%%%%% Fig. 2 %%%%%%%%%%%%%%%%%%%%%%%%%%%%%
\begin{figure}[h]
\begin{center}
\includegraphics[scale=0.4]{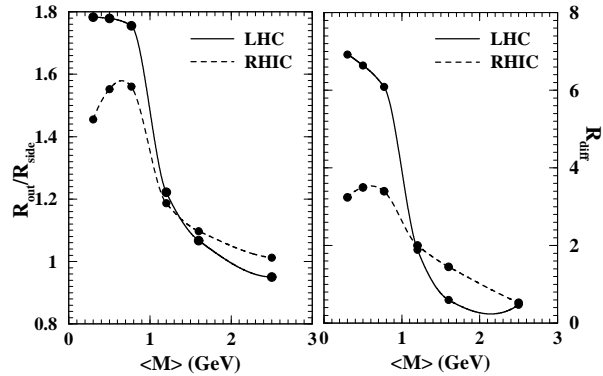}
\caption{The ratio $R_{out}$/$R_{side}$ (left panel) and the difference 
$\sqrt{R_{out}^{2} - R_{side}^{2}}$ (right panel) as a function of  
$\langle M \rangle$ for RHIC (dashed line) and LHC (solid line) energies 
is shown.}
\label{fig2}
\end{center}
\end{figure}
%%%%%%%%%% End of Fig.1 %%%%%%%%%%%%%%%%%%%%%%%%%%%%%

Fig.~\ref{fig2} depicts change of the $R_{out}$/$R_{side}$ and  
$R_{diff} (=\sqrt{R_{out}^{2} - R_{side}^{2}})$ as a function of  
$\langle M \rangle$  
for RHIC as well as LHC energies. Both show a non-monotonic dependence 
on $\langle M \rangle$.  The results reveal that  both the ratio and the
difference of HBT radii for LHC at low $M$ are larger than the corresponding quantities at RHIC. 

Now we indicate below the  experimental challenges in performing
the HBT interferometry with lepton pairs at LHC (for RHIC see~\cite{PM_HBT}).   
We compute the number of events from the luminosity (${\cal L}$),
the $pp$ in-elastic cross section ($\sigma$) and the run time (${\cal T}$) of
the LHC  as:
\begin{equation}
N_{\mathrm event}={{\cal L}}\times \sigma\times {\cal T}
\end{equation}
For ${\cal T}=12$ weeks, ${\cal L}=  
50 \times 10^{27}$/(cm$^2$.sec) and $\sigma$=60 mb
we get $N_{evtent}\sim 2\times 10^{10}$. 
As an example for $\langle M \rangle$ =500 MeV and $k_T=1$ GeV, 
the value of ($dN/d^2k_Tdy$) for Pb+Pb collision at 
$\sqrt{s_{NN}}$=2.76 TeV
is $\sim 0.138\times 10^{-3}$.  
Therefore, the total (for $2\times 10^{10}$ number of events)
differential number of pairs in the above range 
of $k_T$ and $M$ is $\sim 2\times 10^{10}\times 
0.138\times 10^{-3}\sim 2.7\times 10^6$.

Similarly for the $\langle M \rangle$ =1.02 GeV and  $k_T$=1 GeV           
The total (differential) number of pairs is 
$2\times\sim 10^6$. In this domain of $k_T$ and $M$
the number of pairs produced per event is $\sim 10^{-4}$. Therefore,
the probability to get two pairs is
$10^{-8}$, Therefore, roughly $10^{8}$  events will be necessary 
to perform the interferometry  with lepton pairs in this region of $k_T$ and $M$. 

%%%%%%%%%%%%%%%%%%%%%%%%%
\section{Summary}
%%%%%%%%%%%%%%%%%%%%%%
In summary the correlation functions for dilepton pairs has been evaluated 
and the HBT radii have been extracted for both Au+Au collisions at RHIC and Pb+Pb 
collision at LHC energies for different $\langle M\rangle$ windows. We argue that 
the variation of HBT radii with  $M$ for  dilepton pairs can be used as  
an efficient tool to follow the change of the dimension of the evolving system 
with time.  
The quantitative difference in the HBT radii at RHIC and LHC indicate the 
development of larger flow at LHC compared to RHIC.

{\bf Acknowledgment:} We are grateful to Bedangadas Mohanty for 
useful comments.  PM is supported by DAE-BRNS project Sanction No.  2005/21/5-BRNS/2455.

%%%%%%%%%%%%%%%%%%%%%%%%%%%%%%%%%%%%%%%%%%%%%%%%%%%%%%%%%%%

\end{document}